\documentclass[10pt,aps,pre,twocolumn,nofootinbib,superscriptaddress,longbibliography]{revtex4-2}
\usepackage[english]{babel}
\usepackage[utf8x]{inputenc}
\usepackage{amsmath,amsfonts,amssymb,amsthm}
\usepackage{mathpazo}
\usepackage{graphicx}
\usepackage{hyperref}
\usepackage{color}
\usepackage{hyphenat}
\usepackage{natbib}
\usepackage{microtype}
\usepackage{graphicx}
\usepackage{cleveref}
\usepackage{pgfplots}
\usepackage{pgfgantt}
\usepackage[linesnumbered,ruled,vlined]{algorithm2e}
\usepackage{tikz}
\usepackage{mathtools}

\SetKwInput{KwInput}{Input}                
\SetKwInput{KwOutput}{Output}              

\usetikzlibrary{calc}
\pgfplotsset{compat=newest}
\pgfplotsset{plot coordinates/math parser=false}
\newcommand\dif{\mathop{}\!\mathrm{d}}
\newcommand{\brho}{\boldsymbol \rho}
\newcommand{\bsigma}{\boldsymbol \sigma}

\newcommand{\bA}{\mathbf{A}}
\newcommand{\bS}{\mathbf{S}}

\newcommand{\bD}{\mathbf{D}}
\newcommand{\bE}[1]{\operatorname{\mathbb{E}}\left\lbrack #1\right\rbrack}

\newcommand{\bI}{\mathbf{I}}
\newcommand{\bL}{\mathbf{L}}

\newcommand{\bW}{\mathbf{W}}

\newcommand{\Tr}[1]{\operatorname{Tr}\left\lbrack #1\right\rbrack}

\DeclareMathOperator*{\argmin}{arg\,min}

\Crefname{equation}{Eq.}{Eqs.}
\pgfdeclarelayer{background}
\pgfdeclarelayer{foreground}
\pgfsetlayers{background,main,foreground}

\begin{document}
\title{Spectral Entropy via Random Spanning Forests}
\author{Carlo Nicolini\\\url{c.nicolini@ipazia.com}}
\affiliation{Ipazia SpA, Milan, Italy}
\date{\today}
\begin{abstract}
We establish an exact analytic relation between random spanning forests and the heat-kernel partition function. This identity enables estimation of partition functions, energies, and the Von Neumann entropy by Wilson sampling of forests, avoiding costly Laplacian eigendecompositions. We validate inverse-Laplace reconstructions stabilized by a Stieltjes spectral-density regularization on synthetic networks. The approach is scalable and yields local node and edge thermodynamic descriptors.
\end{abstract}
\maketitle

\section{Introduction}
Statistical mechanics on networks and the combinatorics of spanning structures have matured largely in parallel. 
Thermodynamic characterizations based on Laplacian diffusion starts from the partition function of the combinatorial laplacian to encode global observables such as energy, entropy, and spectral moments~\cite{chung1997,dedomenico2016b,nicolini2018}. 
Yet, the combinatorial graph laplacian is also central in random spanning trees and forests describing connectivity through determinantal identities, loop-erased random walks, and Wilson's near-linear-time sampling algorithm~\cite{wilson1996,lyons2017probability,avena2018,avena2018a}. 
While both traditions rely on the Laplacian spectrum, a unified operational framework connecting thermodynamic quantities to combinatorial forest statistics was missing, yet in plain sight~\cite{butzer1968semi}.

In this work we propose an exact bridge between these perspectives by linking the expected root count in random spanning forests to the partition function defined in the spectral entropies framework for complex networks~\cite{dedomenico2016b} by means of inverse Laplace transform.
This implies that thermodynamic observables as energy, von Neumann entropy, and their node resolved counterparts can be recovered from forest statistics without explicit eigendecomposition. 
This is made possible by Wilson sampling~\cite{propp1998,avena2018a}, a method that supplies unbiased Monte Carlo estimators for the expected root count in near-linear time on sparse graphs. 

We demonstrate robust inverse-Laplace reconstructions of spectral partition function using Stieltjes regularized quadrature that tolerates sampling noise.
Numerical experiments on Erd\H{o} s-Rényi, Barabási-Albert, grid, and $k$-regular networks confirm that the forest-based estimators track spectral benchmarks across temperature scales.

This unified framework suggests several directions for future work: extending the forest-heat kernel results to weighted, directed, or temporal networks, or leveraging forest-derived thermodynamic signatures for inference tasks such as network comparison or model selection~\cite{dedomenico2016b}.
These avenues point toward a network information theory grounded simultaneously in combinatorial forests and heat-kernel thermodynamics.

\section{Theory and models}
We summarize here a few definitions which are necessary to make this paper self-contained.
Let us consider undirected graphs $G=(V,E)$ with $|V|=n$ number of nodes and $|E|=m$ number of links. If graphs are weighted, then $W$ denotes the sum of all edge weights, and $w(e)$ indicates the weight of the edge $e$ between nodes $(u,v)$.

In unweighted networs, the binary adjacency matrix is denoted as $\bA=\{a_{ij}\} \in \{0,1\}$ and the (combinatorial) graph Laplacian as $\bL = \bD - \bA$, where $\bD$ is the diagonal matrix of the node degrees.
Similarly, for weighted networks the weighted adjacency matrix is indicated as $\bW = \{ w_{ij} \}$, $w_{ij}\geq 0$ and the weighted Laplacian as $\bL = \bS - \bW$, where $\bS$ is the diagonal matrix of the node strengths.

Notably, the (combinatorial) Laplacian matrix associated with an undirected graph is positive semidefinite, meaning that all its eigenvalues $\lambda_i = \{0,\ldots, 0 \leq \ldots \leq \lambda_n\}$ are nonnegative and real.
Importantly, the multiplicity of the zero eigenvalue counts the number of connected components of the graph $G$.

\subsection{Trees, forests, random walks}
A rooted spanning forest $\phi$ is a directed subgraph of $G$ without cycles, with the same vertex set $V$ of $G$ and such that for each node $u \in V$ there is at most one vertex $v \in V$ where $(u,v)$ is an edge of $\phi$.
The root set $\mathcal{R}(\phi)$ of the forest $\phi$ is the subset of vertices of $V$ for which there is no oriented edge $(u,v)$ in the forest $\phi$.
Being a forest, the connected components of $\phi$ are trees, with edges oriented towards their roots.
The set of all spanning forests $\phi$ of a graph $G$ is denoted by $\mathcal{F}$.

A random spanning forest $\Phi_q$ for $q>0$ is a random variable distributed with law~\cite{avena2018}:
\begin{equation}\label{eq:tree_probability}
	\textrm{Pr}\lbrack \Phi_q=\phi\rbrack = \frac{w(\phi)q^{|\mathcal{R}(\phi)|}} {\chi(q)}.
\end{equation}
Here $w(\phi) = \prod_{e \in \phi} w(e)$ is the weight associated to the forest $\phi$ computed as the product of edge weights.
The number of trees in the forest $\phi$ (or the number of roots) is denoted by the cardinality of the root set $|\mathcal{R}(\phi)|$.
The denominator $\chi(q)$ is a normalization factor. It represents the ``partition function'' of the probability law of Eq.~\ref{eq:tree_probability} and has the meaning of counting all the weighted rooted forests in the set $\mathcal{F}$:
\begin{equation}\label{eq:chi_prob}
	\chi(q) = \sum_{\phi \in \mathcal{F}} w(\phi) q^{|\mathcal{R}(\phi)|}.
\end{equation}
A classical result in graph theory is the Kirchhoff matrix-tree theorem~\cite{lyons2017probability,avena2018,avena2018a}, stating that the number of uniform spanning trees of a graph is specified by the determinant of the minors of the corresponding combinatorial Laplacian $\bL$, or equivalently equals the product of non-zero eigenvalues of $\bL$.
By the matrix-forest theorem~\cite{chebotarev2002forest,chebotarev2006matrix}, the normalization factor $\chi(q)$ in Eq.~\ref{eq:chi_prob} can be written as the determinant of the matrix $q\bI + \bL$:
\begin{equation}\label{eq:chi_partition_function}
	\chi(q) = \det(q\bI + \bL) = \prod_{i=1}^{n}(q+\lambda_i).
\end{equation}

In the case $q=1$, Chebotarev and Shamis~\cite{chebotarev2002forest} found that $\det(\bI + \bL)$ is exactly the (integer) number of rooted spanning forests of the graph $G$.
Equation~\ref{eq:chi_partition_function} is also interpreted as the parametric version of matrix-forests theorem~\cite{agaev2000} to graphs with edges weighted by the factor $1/q$.
In fact, it is simple to see that:
\begin{equation}
	\det(q\bI + \bL) = q^n \det(\bI + \bL/q).
\end{equation}

This result highlights the twofold role of $q$.
On one hand, this parameter is deterministically coupled to the eigenvalues of the combinatorial Laplacian, but at the same time it gives an operative interpretation in terms of combinatorial structures of the graph. It is indeed simple to observe that in the limit $q\to 0$, the random variable $\Phi_q$ describes a random spanning tree on $n$ nodes with exactly $n-1$ links.
In the large $q$ limit instead one obtains a degenerate forest with $n$ trees but no links~\cite{avena2018}.

By standard results on determinantal forest measures~\citep{hough2006determinantal,lyons2017probability,avena2018,avena2018a}, we define the expected number of roots $s(q)$ in the random forest $\Phi_q$:
\begin{equation}\label{eq:s_def}
s(q) := q\frac{d}{dq}\log\chi(q) = \sum_{i=1}^n \frac{q}{q+\lambda_i}.
\end{equation}
The above equation has a dual nature: $s(q)$ it is both a smooth spectral statistic and a direct combinatorial observable sampled efficiently by Wilson's algorithm~\cite{wilson1996,propp1998}.

\subsection{Sampling random spanning forests}
There is an intriguing way to understand the role of $s(q)$ in terms of loop-erased random walks on graphs.
To see this, one can extend the graph $G$ obtaining an augmented graph $G'=(V',E',w')$ with an extra \emph{absorbing} vertex $r$ connected to every node in $V$ to $r$ with directed edges with weight $q$.
Running a loop-erased random walk on the augmented graph defines a series of trajectories, all rooted in $r$.
Then, in order to sample from the distribution in Eq.~\ref{eq:tree_probability}, one only has to remove the root node $r$ from the random spanning tree built on the augmented graph $G'$.

This idea is at the basis of the Wilson algorithm~\cite{wilson1996,propp1998}, a simple yet efficient method to generate random spanning forest with distribution as in Eq.~\ref{eq:tree_probability}.
This algorithm enables to cast a bridge between theory of continuous time random walks on graphs and statistical mechanics via the spectral entropies~\cite{dedomenico2016b,ghavasieh2020statistical} framework.

\begin{figure}[htb]
\centering
\includegraphics[width=0.48\textwidth]{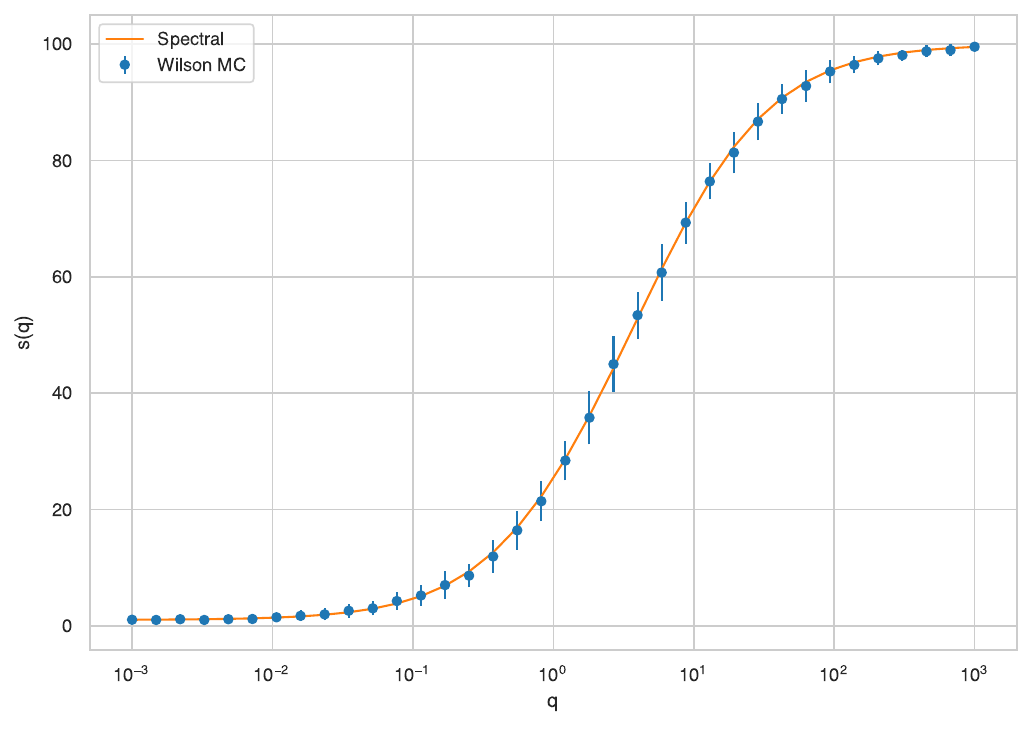}
\caption{Wilson sampling of an Erdos-Renyi graph with 100 nodes and edge probability 0.05. Errorbars are computed as standard deviation from $k=24$ independent runs of Wilson algorithm, solid line is the exact spectral value as in Eq~\ref{eq:s_def}. At low $q$ values the expected number of roots $s(q)$ approaches 1, while at large $q$ it tends to $n=100$.}
\label{fig:s_vs_q}
\end{figure}

Numerically, one can estimate the expected number of roots $s(q)$ from Wilson sampling by generating a total of $k$ i.i.d. forests $\phi_j \sim \Phi_q$ and using the Monte Carlo estimator:
\begin{equation}\label{eq:wilson_montecarlo}
	s(q) = \mathbb{E}\bigl[|\mathcal{R}(\Phi_q)|\bigr] \approx \frac{1}{k} \sum_{j=1}^{k} |\mathcal{R}(\phi_j)|.
\end{equation}
In Figure~\ref{fig:s_vs_q} we show the numerical evidence that Wilson Monte Carlo sampling correctly reconstructs the spectral counterpart of $s(q)$ for an Erdos-Renyi network over a large range of $q$ values.

In the next section we focus on the graph-theoretic aspects of this kind of processes, and explore the tight relation between the combinatorial aspects of random rooted forests and the thermodynamics of heat diffusion in complex networks.

\subsection{Spectral entropies framework}
Building on the analogy between quantum statistical mechanics and network science, the spectral entropies framework proposed by De Domenico and Biamonte~\cite{dedomenico2016b} relies on the definition of the Von Neumann entropy of a network, based on the spectral properties of the network's Laplacian.
This construction leverages the diffusion propagator $e^{-\beta \bL}$, which captures how information spreads through all walks in the network, thereby embedding both local connectivity and global pathways within a single, temperature-dependent information measure.
As a result, the Von Neumann entropy~\cite{dedomenico2016b,wilde2013quantum} defined as 
\begin{equation}\label{eq:vonneumann_entropy}
S(\brho) = - \Tr{\brho \log \brho}
\end{equation}
with $\brho$ being a density matrix defined as the normalized propagator $\brho = \frac{e^{-\beta \bL}}{\Tr{e^{-\beta \bL}}}$, quantifies how rapidly diffusion homogenizes information: regular lattices, modular graphs, and scale-free networks exhibit distinct entropy profiles as a function of $\beta$~\cite{nicolini2020scale}.

Here, the central quantity from which to derive all other thermodynamic observables, is the  partition function $Z(\beta)$, also called \emph{heat-trace}~\cite{xiao2009graph} that takes the form of a trace over the states described by the eigenvalues of $\bL$ weighted by the Boltzmann factor $\beta$:
\begin{equation}\label{eq:z}
	Z(\beta) = \Tr{e^{-\beta\bL}} = \sum_{i=1}^n e^{-\beta \lambda_i}.
\end{equation}
In the next section we will show the deep connection between $s(q)$ and $Z(\beta)$.

\subsection{From random rooted forests to heat trace}
We now give a physical and practical interpretation to the partition function $Z$ in terms of the statistics of the random variable $\Phi_q$ by connecting $s(q)$ and $Z(\beta)$.
Starting from Equation~\ref{eq:z} and of the expected number of roots~\ref{eq:s_def}, we use the elementary identity:
\begin{equation}
\frac{1}{1+a} = \int_0^\infty e^{-(1+a)t}\,\dif t,\qquad a>0,
\end{equation}
replacing $a=\lambda_i/q$ and $t=q\beta$. 
Since $\dif t=q\dif \beta$, the above yields:
\begin{equation}
\frac{q}{q+\lambda_i} = \int_0^\infty q\,e^{-q\beta}e^{-\beta\lambda_i}\,\dif \beta.
\end{equation}
Summing over $i$ and exchanging sum and integral (justified by monotone convergence), we obtain the fundamental equation linking the spectral entropy partition function $Z$ to the combinatorics of random forests described by $s(q)$:
\begin{align}\label{eq:bridge}
s(q) &= \int_0^\infty q\,e^{-q\beta}\left(\sum_{i=1}^n e^{-\beta\lambda_i}\right)\dif \beta \\ 
    &= \int_0^\infty q\,e^{-q\beta} Z(\beta)\,\dif \beta.
\end{align}
Equivalently,
\begin{equation}\label{eq:laplace_transform}
  \frac{s(q)}{q} = \Tr{(q\bI + \bL)^{-1}} = \int_0^\infty e^{-q\beta} Z(\beta)\,\dif \beta
  = \mathcal{L}_\beta[Z](q),
\end{equation}
so that $s(q)/q$ (the resolvent trace) is the Laplace transform in $q$ of the spectral entropy's framework partition function $Z(\beta)$.

This key result establishes that the expected number of roots $s(q)$, a combinatorial quantity obtainable from forest sampling is directly related to the thermodynamic partition function $Z(\beta)$ through a Laplace transform. 
This connection allows us to leverage Laplace transforms to analyze the system's thermodynamics.

\subsection{Network observables from forest statistics}
Beyond global thermodynamic observables such as the heat trace and the Von Neumann entropy, the forest representation naturally suggests local descriptors attached to nodes and edges, which inherit a thermodynamic interpretation through their dependence on the parameter $q$. 
For a given realization of the random forest $\Phi_q$, we define the node-level indicator $X_v(q)=\mathbf{1}\{v\in\mathcal{R}(\Phi_q)\}$, which takes value $1$ if $v$ is a root and $0$ otherwise. 
Its expectation
\begin{equation}
\pi_v(q) = \mathbb{P}\left( v \in \mathcal{R}(\Phi_q) \right) = \bE{X_v(q)}
\end{equation}
plays the role of a local occupation probability for vertex $v$ as a function of $q$. 
By determinantal-forest identities~\cite{lyons2017probability,avena2018}, this probability admits a closed form in terms of the resolvent of the Laplacian,
\begin{equation}\label{eq:root_prob_resolvent}
  \pi_v(q) = q\lbrack (q\bI + \bL)^{-1}\rbrack_{vv}.
\end{equation}
In particular, summing over all vertices recovers the expected root count $s(q)$
\begin{equation}
  \sum_{v\in V} \pi_v(q) = q\,\Tr{(q\bI+\bL)^{-1}} = s(q),
\end{equation}

Equation~\ref{eq:root_prob_resolvent} highlights explicitly how $\pi_v(q)$ encodes spectral information.
Writing the resolvent in its eigenbasis,
\begin{equation}
  (q\bI+\bL)^{-1} = \sum_{i=1}^n \frac{1}{q+\lambda_i}\,u_i u_i^\top,
\end{equation}
with $u_i$ the orthonormal eigenvectors of $\bL$, we obtain
\begin{equation}\label{eq:root_prob_spectral}
  \pi_v(q) = q\sum_{i=1}^n \frac{u_{iv}^2}{q+\lambda_i},
\end{equation}
where $u_{iv}$ denotes the $v$-th component of $u_i$.
Thus the root probability of a node is a superposition of spectral modes, weighted by the kernel $q/(q+\lambda_i)$ and modulated by the local eigenvector weights $u_{iv}^2$. 
Small values of $q$ emphasize contributions from low-lying eigenvalues, probing slow diffusion modes and large-scale connectivity, while large $q$ shift the emphasis toward high-frequency modes. 
From the operational viewpoint, $\pi_v(q)$ can be estimated directly from Wilson sampling by averaging the indicator $X_v(q)$ over independent forests, providing a local thermodynamic descriptor of nodes expressed entirely in terms of forest statistics.
On the other hand, having identified the Laplace transform relation between $s(q)$ and $Z(\beta)$ in Eq.~\ref{eq:laplace_transform}, one can also recover local thermodynamic quantities such as the energy or the Von Neumann entropy at the node level by appropriate integral transforms of $\pi_v(q)$.

A similar construction applies to edges. 
For an undirected edge $e=\{u,v\}$ with weight $w_{uv}$, we define the indicator variable $Y_e(q)=\mathbf{1}\{e\in \Phi_q\}$.
Its expectation value $\pi _v(q)$ is formed as:
\begin{equation}
\theta_e(q) = \mathbb{P}\left( e\in\Phi_q\right) = \bE{Y_e(q)}
\end{equation}
and measures how likely the edge $e$ is to participate in the forest at scale $q$.
Classical results on random forests and effective resistances show that $\theta_e(q)$ can be written in terms of the $s(q)$ as:
\begin{equation}\label{eq:edge_prob_resistance}
  \theta_{\{u,v\}}(q) = w_{uv}\bigl[(q\bI+\bL)^{-1}_{uu} + (q\bI+\bL)^{-1}_{vv} - 2(q\bI+\bL)^{-1}_{uv}\bigr],
\end{equation}
which coincides with the edge weight multiplied by a $q$-dependent effective resistance between $i$ and $j$. 
Expanding again in the spectral basis yields
\begin{equation}\label{eq:edge_prob_spectral}
\theta_{\{i,j\}}(q) = w_{ij}\sum_{k=1}^n \frac{\bigl(u_{ki}-u_{kj}\bigr)^2}{q+\lambda_k},
\end{equation}
so that edges bridging regions where low-frequency eigenvectors vary significantly will tend to have large occupation probability at small $q$, while edges aligned with high-frequency modes become prominent as $q$ increases. 
As for nodes, $\theta_e(q)$ admits a direct Monte Carlo estimator from Wilson sampling by counting how often $e$ appears across independent forests.

These node and edge descriptors, $\pi_v(q)$ and $\theta_e(q)$, constitute local thermodynamic fingerprints of the graph: they interpolate smoothly between different structural regimes as $q$ varies, and they admit a dual interpretation.
On one side they are simple combinatorial observables, accessible by Wilson's algorithm through root and edge frequencies; on the other, they are resolvent-based spectral quantities, explicitly tied to Laplacian eigenvalues and eigenvectors by Eqs.~\ref{eq:root_prob_spectral} and~\ref{eq:edge_prob_spectral}. 
In this sense, the random forest framework not only recovers global thermodynamic functions such as $Z(\beta)$ from forest statistics, but also provides a principled way to assign thermodynamic descriptors at the level of individual nodes and edges, firmly rooted in the spectral geometry of the underlying network.

\section{Numerical experiments}
As described in ~\ref{eq:laplace_transform}, we can recover the partition function $Z(\beta)$ by the inverse Laplace transform of $s(q)/{q}$, which is well-suited problem for the Gaver-Stefehst algorithm~\cite{stehfest1970algorithm} or other numerical Laplace inversion methods~\cite{abate2006unified}.
Indeed, using the spectral definition of $s(q)$ as in Eq.~\ref{eq:s_def}, Gaver-Stefehst inversion already offers a good approximation of $Z(\beta)$.

However, the Gaver-Stehfest algorithm is highly sensitive to numerical noise~\cite{abate2006unified}, which is an important consideration when using Monte Carlo sampling from Wilson's algorithm as in Eq.~\ref{eq:wilson_montecarlo} to estimate $s(q)$.
Different methods can be used to mitigate this issue, the simplest being to increase the number of Wilson samples, but this comes at a computational cost.
In our experiment we evaluate the inverse Laplace transform using a regularized least-squares approach based on Stieltjes transforms, which is more robust to noise and allows to incorporate prior information about the spectral density of the Laplacian, leading to accurate estimates of the $Z$ across a wide range of temperatures.

\begin{figure}[htb]
\centering
\includegraphics[width=0.48\textwidth]{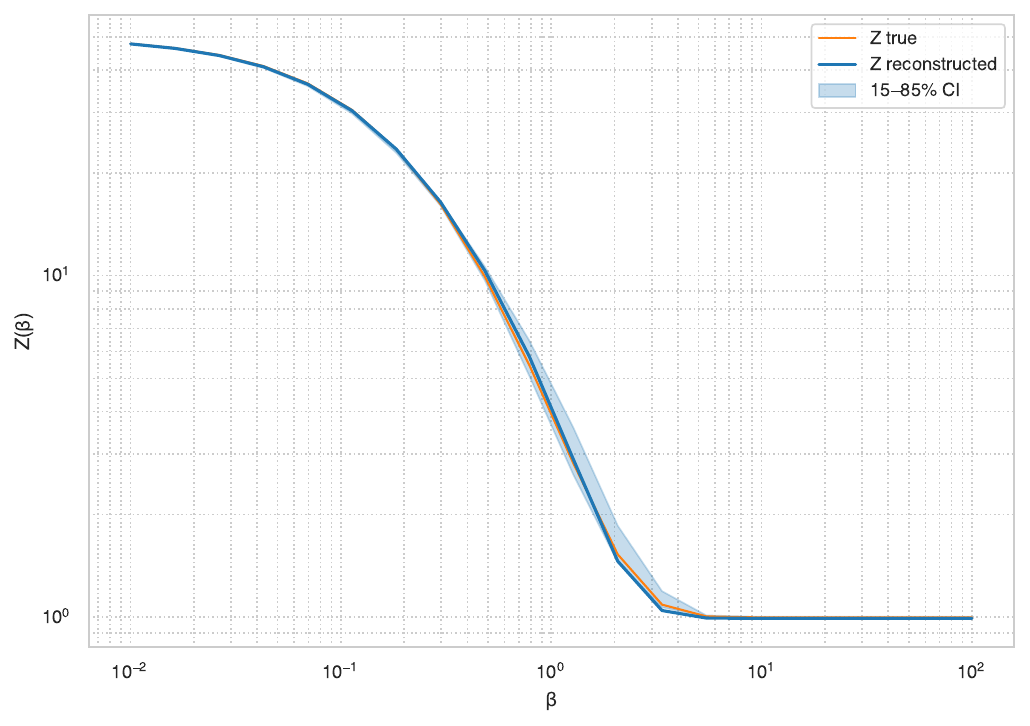}
\caption{Reconstruction of the partition function $Z(\beta)$ from Wilson sampling on an Erd\H{o}s-R\'enyi graph with $n=50$ nodes and edge probability $p=0.1$. The solid blue orange line is the exact spectral value of $Z(\beta)$, the blue solid line is obtained from the Wilson sampling and Stieltjes spectral approximation. Confidence (light blue filled area) are computed as standard deviation on $g(q)$ with $48$ independent Wilson's Monte Carlo samples.}
\label{fig:Z_reconstruction_ER}
\end{figure}

In Figure~\ref{fig:Z_reconstruction_ER} we demonstrate the estimation of the partition function over a range of $\beta$ values for an Erdos-Renyi network, using the spectral estimation based on the Stieltjes transform.
The details of the method are defined in the Appendix~\ref{app:stieltjes}.

\section{Conclusions}
We established exact analytic bridges between random spanning forests and the thermodynamics of diffusion on complex networks. The Laplace transform connects the forest partition function and the expected number of roots to the heat trace, enabling scalable estimation of thermodynamic observables directly from Wilson sampling without computing spectra.

Beyond their conceptual interest, these relations afford practical estimators and diagnostics for large networks.
They provide a principled route to recover spectral moments, to compare ensembles via $s(q)$, and to bound entropy and energy.
Future work will explore weighted and directed settings, numerical Laplace inversion strategies robust to sampling noise, asymptotic predictions for $s(q)$ in random graph models and community-structured graphs.
Altogether, the results open a path toward a rigorous network information theory grounded in forest combinatorics and heat-kernel thermodynamics.

\appendix
\section{Appendix}
\subsection{Code and data availability}
All the code required to run the Wilson algorithm, estimate the $s(q)$ with spectral decomposition method as well as via Monte Carlo sampling are available upon request.
In the appendix we provide the details of the main results of the paper.
\subsection{Wilson algorithm}
We recall the Wilson $q$-algorithm~\cite{avena2018a,propp1998} for sampling random spanning forests with distribution as in Eq.~\ref{eq:tree_probability}.

\begin{algorithm}[htb]
\SetAlgoLined
\DontPrintSemicolon
\KwInput{$G = (V,E,w)$ a weighted undirected graph; optional root weight $q \ge 0$.}
\KwOutput{$F$ a random spanning forest; $\mathcal{R}$ the set of roots.}

\BlankLine
\If{$q > 0$}{
  Add auxiliary root $r$ and edges $(v,r)$ of weight $q$ for all $v \in V$\;
}
Choose root $r$ (auxiliary if $q>0$, otherwise arbitrary)\;
Initialize forest $F \leftarrow \emptyset$, root set $\mathcal{R} \leftarrow \{r\}$\;
Mark all vertices as unvisited\;

\BlankLine
\For{each unvisited vertex $v$ in random order}{
  Perform a random walk from $v$ until it reaches a visited node,
  choosing neighbors with probability $\propto w(u,w)$\;
  Erase loops from the walk to obtain a simple path $P$\;
  Add $P$ to $F$ and mark its vertices as visited\;
}

\BlankLine
\If{$q > 0$}{
  Let $\mathcal{R}$ be all vertices directly linked to the auxiliary root\;
}
\Return{$(F, \mathcal{R})$}
\caption{Random Rooted Forest Sampling (Wilson $q$-algorithm)}
\label{alg:wilson}
\end{algorithm}

\subsection{Numerical inverse Laplace transform via Stieltjes spectral density approximation}\label{app:stieltjes}
We recast the inverse Laplace problem as a numerical quadrature problem based on the estimation of the combinatorial Laplacian spectral density $p_{\bL}(\lambda)$, from which $Z(\beta)$ can then be obtained by a stable forward quadrature. 
Importantly, the estimation is done based on the Monte Carlo estimates of $s(q)/q$ obtained from Wilson sampling.
We start from the definition of the spectral density $p(\lambda)$ as
\begin{equation}
  p(\lambda) = \frac{1}{n}\sum_{i=1}^n \delta(\lambda-\lambda_i),
  \qquad \int_0^\infty p(\lambda)\,d\lambda = 1,
\end{equation}
This let us rewrite both the heat trace and the resolvent in terms of $p(\lambda)$ as:
\begin{equation}
  Z(\beta) = \sum_{i=1}^n e^{-\beta\lambda_i}
           = n \int_0^\infty e^{-\beta\lambda}\,p(\lambda)\,d\lambda,
\end{equation}
and
\begin{equation}\label{eq:stieltjes}
  g(q) = \sum_{i=1}^n \frac{1}{q+\lambda_i}
       = n \int_0^\infty \frac{p(\lambda)}{q+\lambda}\,d\lambda,
\end{equation}
so that $g(q)=s(q)/q$ is recognized as the Stieltjes transform of the spectral density. In practice we do not observe $g(q)$ exactly, but Monte Carlo estimates $\hat g(q_j)$ obtained from Wilson sampling of $\Phi_{q_j}$ as in Eq.~\ref{eq:wilson_montecarlo}.
For each $q_j$ we record the empirical mean $\hat s(q_j)$ of the number of roots and set $\hat g(q_j)=\hat s(q_j)/q_j$, together with an estimated standard error $\sigma_j$.

To turn the inverse problem into a finite-dimensional one, we approximate $p(\lambda)$ on a grid of spectral abscissae $\{\lambda_k\}_{k=1}^{N_\lambda}$ with associated logspaced bin widths $\Delta\lambda_k$.
Thus we represent the density as piecewise constant on the total $N_\lambda$ bins as
\begin{equation}
  p(\lambda) \approx \sum_{k=1}^{N_\lambda} p_k\,\mathbf{1}_{[\lambda_k,\lambda_k+\Delta\lambda_k)}(\lambda),
\end{equation}
and collect the unknown coefficients into a positive vector $\mathbf{p}=(p_1,\ldots,p_{N_\lambda})$.
Inserting this approximation into the Stieltjes transform~\ref{eq:stieltjes} gives, for each sampling point $q_j$,
\begin{equation}
  g(q_j) \approx n \sum_{k=1}^{N_\lambda}
    \frac{p_k \,\Delta\lambda_k}{q_j + \lambda_k}.
\end{equation}
If we gather the values $\hat g(q_j)$ into a data vector $\hat{\mathbf{g}}\in\mathbb{R}^J$, and define the matrix $\bA\in\mathbb{R}^{J\times N_\lambda}$ by
\begin{equation}
  A_{jk} = \frac{n\,\Delta\lambda_k}{q_j+\lambda_k},
\end{equation}
the forward model can be written compactly as
\begin{equation}
  \hat{\mathbf{g}} \approx \bA\,\mathbf{p} + \boldsymbol{\varepsilon},
\end{equation}
where $\boldsymbol{\varepsilon}$ collects the Monte Carlo fluctuations induced by Wilson sampling.

Because different $q_j$ yield estimates with different uncertainty, we weight the residuals by their inverse variances. Denoting by $\bsigma=(\sigma_1,\ldots,\sigma_J)^\top$ the standard errors of $\hat g(q_j)$, we introduce the diagonal weight matrix
\begin{equation}
\bW = \mathrm{diag}\!\left(\frac{1}{\sigma_1^2},\ldots,\frac{1}{\sigma_J^2}\right),
\end{equation}
so that $\|\bW^{1/2}(\bA\mathbf{p}-\hat{\mathbf{g}})\|_2^2$ is a weighted least-squares data term.
The spectral density must also satisfy the normalization constraint
\begin{equation}
  \sum_{k=1}^{N_\lambda} \Delta\lambda_k\,p_k \approx 1,
\end{equation}
which in discretized form ensures that $p(\lambda)$ integrates to one. 
Rather than enforcing this as a hard constraint, we include it as a quadratic penalty with strength $\gamma_{\mathrm{mass}}>0$, which adds the term
\begin{equation}
\gamma_{\mathrm{mass}}\left(\sum_{k=1}^{N_\lambda} \Delta\lambda_k p_k - 1\right)^2
\end{equation}
to the objective. 

Finally, to regularize the ill-posed inversion and suppress spurious oscillations in $\mathbf{p}$ induced by noise, we penalize curvature of the reconstructed density by a discrete second-derivative operator. 
Concretely, we let $\bD\in\mathbb{R}^{(N_\lambda-2)\times N_\lambda}$ be the second-difference matrix acting as
\begin{equation}
  (\bD\mathbf{p})_\ell = p_\ell - 2p_{\ell+1} + p_{\ell+2}, \qquad \ell=1,\ldots,N_\lambda-2,
\end{equation}
and introduce a Tikhonov smoothness term $\tau_{\mathrm{smooth}}\|\bD\mathbf{p}\|_2^2$ with hyperparameter $\tau_{\mathrm{smooth}}>0$ controlling the amount of regularization.

Putting these ingredients together, the reconstruction of the spectral density reduces to the nonnegative Tikhonov minimization problem:
\begin{align}
\mathbf{p}^\star
= \argmin_{\mathbf{p}\ge 0}\;
    &\left\|\bW^{1/2}\left(\bA\mathbf{p}-\hat{\mathbf{g}}\right)\right\|_2^2 \\
    & + \gamma_{\mathrm{mass}}\left(\sum_{k=1}^{N_\lambda}\Delta\lambda_k p_k - 1\right)^2 \\
    & + \tau_{\mathrm{smooth}}\|\bD\mathbf{p}\|_2^2 .
\end{align}
The partition function is finally recovered by a stable forward quadrature as
\begin{equation}
  Z(\beta) \approx
  n \sum_{k=1}^{N_\lambda} p_k^\star\, e^{-\beta \lambda_k}\,\Delta\lambda_k,
\end{equation}
which amounts to the inverse Laplace transform of $g(q)$ as defined in Eq.~\ref{eq:stieltjes}.

\bibliographystyle{plain}
\bibliography{biblio}

\end{document}